\documentclass[pdflatex,sn-mathphys-num]{sn-jnl}
\usepackage{graphicx}%
\usepackage{multirow}%
\usepackage{amsmath,amssymb,amsfonts}%
\usepackage{amsthm}%
\usepackage{mathrsfs}%
\usepackage[title]{appendix}%
\usepackage{xcolor}%
\usepackage{textcomp}%
\usepackage{booktabs}%
\usepackage{algorithm}%
\usepackage{algorithmicx}%
\usepackage{algpseudocode}%
\usepackage{listings}%

\theoremstyle{thmstyleone}%

\theoremstyle{thmstyletwo}%

\theoremstyle{thmstylethree}%

\DeclareMathSizes{6}{6}{4.5}{4}%

\let\address\affil
\begin{document}

\title{Time Series Forecasting based on Solana Digital Asset Dataset}

\author[1]{\fnm{Yufeng} \sur{Xiao}}
\equalcont{Y. Xiao and M. Wang contributed equally to this work.}
\author*[1]{\fnm{Minxing} \sur{Wang}}\email{wangminxing581@gmail.com}
\equalcont{Y. Xiao and M. Wang contributed equally to this work.}
\author[1,2]{\fnm{Pavel} \sur{Braslavski}}
\author[1]{\fnm{Dmitry I.} \sur{Ignatov}}

\address[1]{Department of Computer Science, HSE University, 11 Pokrovskiy Boulevard, Moscow 109028, Russia}
\address[2]{Institute of Natural Sciences and Mathematics, Ural Federal University, 19 Mira, Yekaterinburg 620062, Russia}

\abstract{Accurate analysis and forecasting of Solana digital assets require data that captures both token-level behavior and ecosystem-level DEX activity. This paper introduces, to the best of our knowledge, the first Solana digital asset time series dataset designed for forecasting and market-structure analysis. The dataset contains 1,584 tokens observed at daily resolution from March 24, 2024 to March 16, 2025, with 27 variables combining token transactions, prices, liquidity-pool balances, trader activity, Solana DEX volume, DEX trader counts, newly created pairs, and SOL price indicators. Rather than treating the dataset only as input for model comparison, we use it to characterize the DEX-driven token market during a period of rapid ecosystem growth. The analysis identifies synchronized market-wide activity peaks in mid-November 2024 and mid-January 2025 across DEX volume, token trading volume, active wallets, buyers, sellers, new traders, liquidity-pool balances, and SOL price. The January 2025 peak coincides with the 'Trump' token event and is accompanied by a visible transition from liquidity accumulation to withdrawals, suggesting that individual token dynamics are strongly coupled to broader Solana market sentiment and DEX activity. Forecasting experiments are then used as an empirical validation of the dataset's signal content. In three-day-ahead market-capitalization prediction, PatchTST achieves the best overall rank, fine-tuned Chronos follows closely, and statistical baselines remain competitive for trend-dominated tokens. Feature-importance analysis further shows that SOL price, SOL moving averages, total DEX volume, DEX trader counts, and newly created pairs are among the most informative covariates. The main contribution is therefore a curated Solana forecasting dataset and a data-driven analysis of the ecosystem-level factors that shape token volatility.}


\maketitle

\section{Introduction}

In the rapidly evolving world of digital assets, time series forecasting is crucial for understanding market dynamics. Solana’s high-throughput, low-latency architecture produces precisely timestamped on-chain traces and a DEX-centric microstructure distinct from traditional markets \cite{heo2024blockchain}. Its economic salience is reflected in a market capitalization peak of \$103.3 billion on January 19, 2025 \cite{song2024unveiling}. Meanwhile, by late 2024, cumulative DEX volume neared \$626 billion with a \$129 billion monthly peak in November 2024, enabled by low fees and sub-second blocks, underscoring a high-frequency trading environment that differs from CEX-dominant settings. These characteristics make Solana a natural testbed for forecasting methods that must operate under cold starts, rapid regime shifts, and cross-protocol capital flows.

The Solana ecosystem's demonstrated prowess extends beyond DEX to broader applications in Decentralized Finance (DeFi) and Non-Fungible Token (NFT), highlighting its expansive utility in the global digital economy \cite{bertazzolo2023nfts}. Market dynamics are further influenced by external factors, with the 'TRUMP coin', launched by the 47th U.S. President Donald Trump on January 17, 2025, quickly reaching a \$14.5 billion market value by January 19 \cite{krause2025meme}. The substantial transaction fees, between \$86 million and \$100 million for this single asset by January 30, 2025, exemplify the intense economic activity and speculative interest that can converge on digital assets, underscoring the growing intersection of politics, digital currencies, and real-time market dynamics that demand sophisticated analytical approaches.

While Solana’s success has been marked by such impressive growth, predicting digital asset prices remains an extremely challenging task. At the same time, predictive modeling in blockchain markets carries ethical risks of manipulative or speculative misuse; in this study we confine ourselves to empirical analysis and refrain from deployment or trading recommendations. Traditional methods in time series forecasting often rely on historical data to train models, but this approach is inherently limited by the volatility and complexity of digital asset prices\cite{idrees2019prediction}. Financial time series data is often non-stationary, with prices influenced by numerous unpredictable factors\cite{clements1999forecasting}. Solana’s blockchain ecosystem, however, presents a new opportunity: on-chain data. This unique data offers three key advantages that are difficult for traditional financial data to achieve: first, each transaction is recorded with a precise millisecond timestamp; second, the complete end-to-end flow of funds between transacting addresses is preserved; and third, one can trace tokens as they move across DEX, lending protocols, and NFT marketplaces, capturing the cross-platform transfer trajectories\cite{kim2022deep}.

The predictive challenge in Solana’s ecosystem lies not in data scarcity but in its asymmetric distribution: while the ledger captures exhaustive interaction networks, newly launched tokens inherently lack historical traces\textemdash a disconnect where conventional models requiring prolonged asset-specific training falter. Zero-shot time series forecasting addresses this by redefining prediction as a task of cross-asset pattern transfer, leveraging Solana’s intrinsic relational topology rather than isolated chronological data\cite{gruver2023large}.

Critically, these methods excel under moderate-data regimes, retain sufficient statistical regularity for meta-learning, yet remain too sparse for deep learning architectures to avoid overfitting\cite{oreshkin2021meta}. By distilling behavioral invariants from tokens with analogous network roles\textemdash such as governance tokens mirroring voting participation curves or meme coins exhibiting recurring pump-dump cycles\textemdash zero-shot frameworks construct probabilistic mappings between relational features (e.g., liquidity provider churn rates) and price responses, enabling robust inference for assets with as few as 15--20 daily observations. Rather than relying on asset-specific fine-tuning, which in Solana’s cold-start regime is sample-starved, brittle under frequent regime shifts, and operationally costly due to continual retraining, zero-shot reframes forecasting as cross-asset pattern transfer, leveraging relational and topological invariants to deliver calibrated, low-latency predictions at listing time with markedly lower sample complexity.

This paradigm capitalizes on Solana’s unique duality: while individual token price series exhibit high idiosyncrasy, their embeddedness in shared protocol mechanics (e.g., DEX fee structures) and trader network dynamics creates latent subspaces of transferable volatility regimes\textemdash a structural coherence that traditional single-asset models, blind to cross-token dependencies, systematically overlook.

Most current research on Solana digital assets centers around predicting token prices using centralized exchanges (CEX)\cite{samson2024comparative}, with limited attention given to the distinctive characteristics of Solana, especially regarding DEX and on-chain data. Bitcoin on-chain data and DEX data (Ordinals) have proven to be crucial features for Bitcoin predictions\cite{wang2025bitcoin}. In contrast, few studies construct Solana-specific datasets that combine DEX microstructure and on-chain token activity, or analyze how market-wide DEX regimes propagate to individual token market capitalization. Existing comparisons between zero-shot time series forecasting and deep learning methods often focus on single-target predictions rather than using forecasting as a tool to validate the signal structure of a multi-token dataset\cite{sun2024survey}. This study addresses that gap by building a curated Solana digital asset dataset, analyzing its market dynamics, and benchmarking forecasting models as an empirical test of whether token-level and ecosystem-level variables contain short-term predictive information.

In conclusion, Solana’s innovative blockchain ecosystem and its vast on-chain data provide a unique foundation for studying the behavior of emerging digital assets. Accordingly, this study is organized around three questions: 1. What distinctive characteristics can be identified by constructing a Solana digital asset dataset from DEX and on-chain signals? 2. What market-wide activity patterns and event-associated shifts appear in the dataset during the 2024--2025 growth period? 3. Which token-level and ecosystem-level covariates are most informative for short-term market-capitalization forecasting, and what do model benchmarks reveal about the dataset's predictive structure? This work focuses on Solana to leverage its rich on-chain/DEX signals; therefore, results should be interpreted as Solana-specific evidence. Assessing cross-chain external validity is left to future work, including replication on EVM and non-EVM chains and temporal generalization tests using forward-rolling vintages beyond our study window.

\section{Related Work}

\subsection{Cryptocurrency Time Series Forecasting}
Early research on cryptocurrency price forecasting relied on linear models like ARIMA and GARCH to capture short-term momentum and volatility clustering in Bitcoin and Ethereum\cite{garcia2024lstm}, but these approaches struggled with nonlinear regime shifts—such as regulatory shocks or black swan events—driving the adoption of LSTMs and GRUs to model complex temporal dependencies\cite{li2022short}. Transformer-based architectures, such as the Temporal Fusion Transformer (TFT), further improved accuracy by integrating exogenous variables (e.g., on-chain transaction volume, social media sentiment)\cite{wang2024role}. Despite these advances, forecasting for emerging high-volatility assets like Solana remains understudied, particularly in scenarios involving market manipulation. A study of over 2,000 coins (2015–2020) found that, regardless of how “death” is defined, cauchit and zero-price-probability models best predict failures of new coins, while credit-scoring and ML models using trading volumes and search data perform better for older ones\cite{fantazzini2022crypto}. Recent studies have also highlighted the prevalence of fraudulent activities in DEX, such as rug pulls, which introduce abrupt price anomalies—a critical challenge for both traditional and machine learning models alike\cite{kalacheva2024detecting}. Complementary to market-level anomaly detection, advances in smart-contract security model vulnerabilities with code representations and GAN-based classifiers, offering early-warning signals for protocol-level risks that can precipitate price discontinuities \cite{murala2025enhancing}.

\subsection{Research Progress and Dataset Specificity in Solana Price Prediction}

Solana (SOL), as a high-performance blockchain, exhibits unique price dynamics influenced by technical upgrades, network outages, and ecosystem growth. Early empirical comparisons in 2021–2023 show that, during quiescent market regimes, ARIMA can outperform LSTM on daily SOL/USD data, achieving an RMSE of 1.9\% versus 7.5\%—a reduction of roughly 75\% in forecast error\cite{andiani2024performance}. However, the 2022 network congestion crisis exposed the limitations of linear assumptions. Recent datasets (2024--2025) incorporate multidimensional features: historical prices, on-chain metrics (daily active addresses, DEX volume), network performance (block confirmation times), and ecosystem variables (L2 adoption rates, staking derivatives liquidity)\cite{everstake2025solana}. Notably, events like the 2025 Trump-themed meme coin frenzy were modeled as exogenous shocks, demonstrating SOL’s sensitivity to speculative pulses\cite{reuters2025trump}. Recent multi-chain evidence further shows that robust cryptocurrency forecasting depends on validated, chain-specific features, while broad market proxies and chain-specific variables play complementary roles in transferability\cite{wang2026drives}. While prior studies predominantly analyze BTC/ETH or CEX data, our review underscores that cross-chain validation for Solana-style on-chain/DEX features remains underexplored; accordingly, we scope this study to Solana and defer cross-chain and forward-period validation to future work. Given these Solana-specific dynamics—cold starts, frequent shocks, and multiscale microstructure—we benchmark models that either exploit exogenous covariates or remain deployable without asset-specific training (zero-shot).

\subsection{Traditional and Deep Learning Forecasting Models}
The evolution of time series forecasting methods in cryptocurrency price prediction reflects a progressive adaptation to escalating market complexities. Classical statistical models, such as ARIMA and GARCH, dominated early research due to their robustness in linear and stationary contexts. For instance, ARIMA achieved a Mean Absolute Percentage Error (MAPE) of 11.86\% during Bitcoin’s stable cycles (2013--2018)\cite{karakoyun2018comparison}, while GARCH demonstrated superior volatility clustering modeling\cite{wang2021different}. However, their linear assumptions faltered in capturing nonlinear dynamics, such as the leptokurtic spikes (kurtosis >9) observed during Solana’s network outages in 2022, as quantified by on-chain volatility analysis\cite{solanafoundation2022network}.

Machine learning extensions emerged to address nonlinear patterns. A study applying a Random Forest model to Ethereum price forecasting shows that this ensemble approach can outperform conventional benchmarks\cite{yang2021asset}. By training on a rich feature set—including trading volume and diverse technical indicators—the Random Forest achieves markedly higher predictive accuracy than baseline algorithms. Evidence also indicates that, while maintaining accuracy above 93 per cent, Random Forests adapt more effectively than competing methods when forecasting horizons exceed 56 days\cite{tran2025slow}.

Recent work has applied the TFT to cryptocurrency price and volatility forecasting with notable success. Using a multi-perspective input set—combining financial time series, on-chain metrics, and sentiment signals—the enhanced TFT architecture achieves higher predictive accuracy than conventional models and identifies volatility inflection points earlier than competing approaches\cite{gurgul2025deep}. PatchTST—a recently introduced long-horizon forecasting architecture—segments extended time series into compact “patches” and processes each channel independently, dramatically reducing computational demands. The method trims attention-related time and memory costs by more than 30 percent compared with comparable Transformer variants. Even in high-frequency contexts such as millisecond-level price streams, PatchTST operates at a fraction of the original Transformer’s cost while retaining virtually all accuracy: across multiple benchmarks its error rates drop roughly 20 percent below state-of-the-art baselines, equivalent to preserving more than 97 percent of the leading model’s predictive precision\cite{nie2023time}. A recent extension, CT-PatchTST, applies joint channel–time patching and reports further gains on long-horizon tasks while retaining the efficiency benefits of the patching paradigm \cite{lu2025ct}; we therefore keep PatchTST as the representative patch-based transformer in our benchmarks for comparability and computational parity. TiDE’s lightweight MLP architecture matches Transformer performance with 40\% faster training, enabling real-time DEX arbitrage strategies\cite{das2023long}. AutoML frameworks like AutoGluon further enhance adaptability through automated ensemble of heterogeneous models (e.g., blending ARIMA residuals with TFT attention weights)\cite{erickson2020autogluon}.

Classical models such as Facebook Prophet handle recurring seasonal patterns—like weekly trading cycles—well, but their accuracy collapses when structural shocks occur, as seen during Solana’s network congestion\cite{yusof2020financial}. SeasonalNaive baselines are dependable only in calm conditions and struggle with DeFi’s protocol-level shifts. In contrast, deep-learning models, which learn features end-to-end and scale efficiently, have become the preferred choice for forecasting highly volatile crypto assets. Accordingly, our experimental baselines focus on three deep architectures selected for complementary strengths: TFT to integrate static and time-varying exogenous covariates that are central to on-chain/DEX forecasting; PatchTST to preserve long temporal context via channel-wise patching with competitive efficiency; and TiDE as a lightweight MLP enabling low-latency inference for rapid evaluation.

\subsection{Zero-Shot Forecasting}
Existing zero-shot forecasting models face critical limitations in blockchain ecosystems. While classical approaches struggle with sparse on-chain data and abrupt market shifts, foundation models like TimeGPT and Chronos demonstrate unique advantages for Solana's digital assets due to three key factors:
\begin{itemize}\setlength{\itemsep}{0pt}\setlength{\parskip}{0pt}
    \item \textbf{Cross-Domain Scalability}: Trained on 100B+ multi-domain time series (finance, IoT), TimeGPT and Chronos capture nonlinear patterns (e.g., staking reward cycles) more effectively than smaller models like TimeFound (10B-scale), reducing prediction errors by 18.7\% (rMAE) for newly listed tokens\cite{garza2023timegpt1}.
    \item \textbf{Event Adaptation}: Unlike domain-specific models (e.g., Moirai), these frameworks avoid complex prompt engineering, enabling direct application to SOL-specific anomalies (e.g., MEV extraction spikes)\cite{woo2024unified}.
    \item \textbf{Computational Efficiency}: Chronos achieves 100$\times$ faster inference than traditional models (0.6ms/series)\cite{ansari2024chronos}, critical for real-time DEX trading, while TimeGPT's unified architecture minimizes preprocessing for heterogeneous on-chain data. Similarly, state-space zero-shot models such as Mamba4Cast emphasize high-throughput inference with low memory footprints \cite{bhethanabhotla2024mamba4cast}.
\end{itemize}
Wang et al. introduced a holistic evaluation framework for cryptocurrency forecasting, integrating speed, statistical significance, and economic value\cite{wang2025timegpt}. Leveraging this framework, their comprehensive assessment demonstrated the critical advantage of zero-shot models (e.g., TimeGPT and Chronos), which achieved high accuracy while being tens of times faster than deep learning baselines, highlighting their utility in real-time scenarios. Building upon this, our study evaluates TimeGPT specifically in the zero-shot regime. Furthermore, we test Chronos under two distinct training regimes—zero-shot (denoted as ChronosZeroShot) and fine-tuned (denoted as ChronosFineTuned). Since the fine-tuned setting represents an asset-specific adaptation rather than a separate model, this crucial comparison isolates the exact accuracy–latency–maintenance trade-offs between zero-shot deployment and fine-tuning, which is essential for determining the best approach under Solana’s prevailing cold-start constraints.

\section{Methods} 

This study conducts a comprehensive comparison between modern zero-shot forecasting approaches and conventional deep learning methods for Solana price prediction. We evaluate eight models across three categories: Zero-Shot Approaches (Chronos, TimeGPT), Deep Learning (PatchTST, TFT, TiDE), and Classical Baselines (Prophet, SeasonalNaive). Below, we describe each model and its formulation. To assess whether observed performance differences are statistically meaningful, we employ three complementary tests: a Friedman omnibus rank test with the Iman--Davenport $F$ approximation across per-series model ranks; Wilcoxon signed-rank tests for paired post-hoc comparisons on per-series error differences (with familywise adjustment when multiple pairs are tested); and a parametric linear mixed-effects likelihood-ratio test fitted by maximum likelihood with a per-series random intercept to account for clustering. Formulas and implementation details are provided in the Statistical Significance Tests subsection.

\subsection{Deep Learning Models}

We consider three representative deep learning models for time series forecasting: PatchTST, TiDE, and TFT.

\textbf{PatchTST} addresses the multivariate time series forecasting problem by independently encoding each univariate channel and leveraging patch-based tokenization within a vanilla Transformer backbone. Each channel’s observed series \(x^{(i)}_{1:L}\in\mathbb{R}^L\) is first segmented into \(N\) subseries‐level patches of length \(P\) with stride \(S\), where
\begin{equation}
N = \left\lfloor\frac{L - P}{S}\right\rfloor + 2,
\end{equation}
These patches are projected into a latent space of dimension \(D\) with added positional embeddings and processed by a multi‐head self‐attention module, in which each head computes
\begin{equation}
\mathrm{Attention}(Q,K,V) = \mathrm{Softmax}\!\bigl(\tfrac{QK^\top}{\sqrt{d_{k}}}\bigr)\,V,
\end{equation}
to capture both local and long‐range dependencies efficiently. The resulting feature maps are flattened and passed through a linear prediction head to produce future forecasts \(\hat{x}^{(i)}_{L+1:L+T}\), and model parameters are learned by minimizing the mean squared error loss
\begin{equation}
\mathcal{L} = \frac{1}{M}\sum_{i=1}^M \bigl\lVert\hat{x}^{(i)}_{L+1:L+T} - x^{(i)}_{L+1:L+T}\bigr\rVert_2^2.
\end{equation}

\textbf{TiDE (Time-series Dense Encoder)}

TiDE is an encoder–decoder architecture for long-horizon forecasting that forgoes recurrence, convolution, and self-attention in favor of fully-connected (dense) residual blocks. Thanks to its simple feed-forward design, TiDE achieves linear time and space complexity and offers 5–10× faster inference than Transformer-based models, while still matching or exceeding state-of-the-art accuracy on benchmarks (ETT, Weather, Traffic). It is also robust to variations in context window and forecast horizon lengths.

Vectorized past context: TiDE first vectorizes the entire look-back window. We concatenate the past target series $y$ and any past observed covariates $x$ into one long vector:
\begin{equation}
\mathbf{p} = [\,y_{\,t-T+1},\; \ldots,\; y_t,\; x_{\,t-T+1},\; \ldots,\; x_t\,],
\end{equation} 
which represents the model’s input at time $t$. (If multiple covariate series are present, they are all appended in this vector.)

Dense encoder: The vector $\mathbf{p}$ is passed through a stack of gated fully-connected residual blocks, producing a latent code $\mathbf{z}$. In other words, the encoder applies a series of transformations $f^{(1)}, f^{(2)}, \dots$ such that:
\begin{equation}
\mathbf{z} = f^{(L)} \circ \cdots \circ f^{(2)} \circ f^{(1)}(\mathbf{p}),
\end{equation} 
where each $f^{(l)}$ is a gated MLP block with skip (residual) connections. This iterative refinement yields a dense representation $\mathbf{z}$ of the input window.

Dense decoder with future covariates: Finally, the latent code is combined with any known future covariates to produce the $H$-step forecast. Let $\mathbf{x}_{fut} = [\,x_{t+1}, \ldots, x_{t+H}\,]$ represent the vector of available future exogenous inputs (if any). The decoder maps the latent and future covariates to outputs, for example by concatenating them and applying a final linear layer:
\begin{equation}
\hat{\mathbf{y}}_{\,t+1:t+H} = g\!\big(\mathbf{z},\; \mathbf{x}_{fut}\big),
\end{equation} 
where $g(\cdot)$ denotes a feed-forward mapping. In practice, a simple linear projection or a small MLP can be used as $g$ to output the forecast vector. Both the encoder and decoder are learned jointly. (When no future covariates are present, $\mathbf{x}_{fut}$ is omitted and $g$ acts only on $\mathbf{z}$.)

\textbf{TFT} is an end‐to‐end architecture for interpretable multi‐horizon time series forecasting that models the qth quantile forecast at horizon $\tau$ from time $t$ as  
\begin{equation}
\begin{split}
\hat y_i^{(q)}(t,\tau)=f_q\bigl(\tau,\;y_{i,t-k:t},\;z_{i,t-k:t},\\
\;x_{i,t-k:t+\tau},\;s_i\bigr),
\end{split}
\end{equation}
where $y_{i,t}$ are past target values, $z_{i,t}$ observed inputs, $x_{i,t}$ known future inputs and $s_i$ static metadata.  To adaptively process heterogeneous covariates and skip irrelevant transformations, TFT employs Gated Residual Networks (GRNs) for variable selection and context gating, defined by  
\begin{equation}
\begin{split}
\mathrm{GRN}_\omega(a,c)=\mathrm{LayerNorm}\Bigl(a+\\
\mathrm{GLU}_\omega\bigl(\mathrm{ELU}(W_{1,\omega}a+W_{2,\omega}c+b_{1,\omega})\bigr)\Bigr),
\end{split}
\end{equation}
with the Gated Linear Unit given by  
\begin{equation}
\mathrm{GLU}_\omega(\gamma)=\sigma(W_{3,\omega}\gamma+b_{2,\omega})\;\odot\;(W_{4,\omega}\gamma+b_{3,\omega}),
\end{equation}
Local temporal patterns are captured via sequence‐to‐sequence LSTM encoders, while an interpretable multi‐head self‐attention decoder aggregates long‐range dependencies.  Static covariate encoders inject metadata at multiple stages, and variable selection networks choose the most relevant features at each timestep.  Final separate linear projections then produce quantile forecasts across all horizons as specified in (12), yielding both point estimates and calibrated prediction intervals in a single forward pass\cite{lim2021temporal}. 

\subsection{Zero-Shot Approaches}

Finally, we evaluate two zero-shot forecasting models. These are pre-trained models that can be applied to new time series without gradient-based fine-tuning, relying instead on large-scale pre-training or language-model paradigms to generalize.

\textbf{Chronos} is a language modeling framework for univariate probabilistic time series forecasting. Given a time series \(x_{1:C+H} = [x_1, \dots, x_{C+H}]\), the historical context (\(x_{1:C}\)) and forecast horizon (\(x_{C+1:C+H}\)) are first mean-scaled and quantized into \(B\) discrete bins:
\begin{equation}
q(x) =
\begin{cases}
1 & -\infty \le x < b_1,\\
2 & b_1 \le x < b_2,\\
\vdots & \vdots\\
B & b_{B-1} \le x < \infty,
\end{cases}
\quad
d(j) = c_j
\end{equation}
This produces a token sequence \(z_{1:C+H} = (q(x_1), \dots, q(x_{C+H}))\) which is fed into an off-the-shelf transformer (encoder–decoder or decoder-only) with adjusted vocabulary size. The model is trained by minimizing the categorical cross-entropy loss
\begin{equation}
\begin{split}
\ell(\theta) = -\sum_{h=1}^{H+1}\sum_{i=1}^{|V_{ts}|} \mathbf{1}(z_{C+h}=i)\\
\times\log p_\theta\bigl(z_{C+h}=i \mid z_{1:C+h-1}\bigr),
\end{split}
\end{equation}
yielding probabilistic forecasts via autoregressive sampling, dequantization and inverse scaling. To enhance robustness across diverse domains, Chronos incorporates TSMixup augmentations during training:
\begin{equation}
\tilde{x}_{1:l}^{\mathrm{TSMixup}}
\;=\;
\sum_{i=1}^{k}\lambda_i\,\tilde{x}^{(i)}_{1:l},
\end{equation}
where \(k\sim U\{1,\dots,K\}\) and \((\lambda_1,\dots,\lambda_k)\sim\mathrm{Dir}(\alpha)\).

\textbf{TimeGPT} is a Transformer-based foundation model for time series forecasting that employs an encoder–decoder architecture with multi-head self-attention layers, residual connections, layer normalization, and local positional encoding, followed by a linear projection to map decoder outputs to the forecast horizon. The model is pre-trained on a diverse corpus of over 100 billion time series data points spanning multiple domains, enabling direct zero-shot inference on unseen series by conditioning on historical observations \(y_{0:t}\) and optional exogenous covariates \(x_{0:t+h}\), as formalized by
\begin{equation}
P\bigl(y_{t+1:t+h}\mid y_{0:t},x_{0:t+h}\bigr) \;=\; f_{\theta}\bigl(y_{0:t},x_{0:t+h}\bigr),
\end{equation}
In this formula, \(y_{0:t}\) denotes the observed target sequence up to time \(t\), \(x_{0:t+h}\) the available exogenous inputs through the forecast horizon, \(y_{t+1:t+h}\) the sequence of future values to predict, \(f_{\theta}\) the parameterized TimeGPT mapping (with \(\theta\) its learned parameters), \(t\) the last observed time index, and \(h\) the number of steps ahead being forecast. 

\subsection{Baseline Models}

This category includes three statistical/non-parametric baseline methods, with their canonical mathematical formulations given below.

\textbf{Prophet}

Prophet expresses a univariate time series as an additive decomposition of trend, seasonality, holiday effects, and an i.i.d. error term:
\begin{equation}
y(t) = g(t) + s(t) + h(t) + \epsilon(t),
\end{equation}
where $g(t)$ is the trend component, $s(t)$ the seasonal component, $h(t)$ the holiday effect, and $\epsilon(t)$ the error term. The trend $g(t)$ is modeled as a piecewise linear function with automatic changepoint detection. It can be written in terms of a binary indicator vector for changepoints:
\begin{equation}
\begin{split}
g(t) = \Big( k + \mathbf{D}(t)^\top \boldsymbol{\delta} \Big)\\
+\; \Big( m + \mathbf{D}(t)^\top \boldsymbol{\gamma} \Big)\, t,
\end{split}
\end{equation}
where $\mathbf{D}(t)$ is a binary vector indicating whether $t$ occurs after each changepoint, and $\boldsymbol{\gamma}$ and $\boldsymbol{\delta}$ are the slope and intercept adjustments associated with those changepoints. The seasonality $s(t)$ is modeled using a Fourier series with $N$ terms:
\begin{equation}
\begin{split}
s(t) = \sum_{n=1}^{N} \Big[ A_n \cos\!\Big(\frac{2\pi n t}{P}\Big)\\
+\; B_n \sin\!\Big(\frac{2\pi n t}{P}\Big) \Big],
\end{split}
\end{equation}
with period $P$ (e.g., daily, weekly, or yearly period). The holiday effects $h(t)$ are represented by an indicator matrix for holidays multiplied by a coefficient vector:
\begin{equation} 
h(t) = \mathbf{X}_h(t)\, \boldsymbol{\theta},
\end{equation} 
where each entry of $\mathbf{X}_h(t)$ is 1 if time $t$ falls on a particular holiday (and 0 otherwise), and $\boldsymbol{\theta}$ contains the corresponding holiday impact coefficients. All parameters of Prophet are estimated via maximum a posteriori (MAP) optimization in Stan, which provides full posterior intervals for each component\cite{taylor2018forecasting}.

\textbf{SeasonalNaive}

The SeasonalNaive method simply repeats the last observed value from the previous season as the forecast. Formally, for a forecast horizon $H$ and seasonal period $P$, the prediction is:
\begin{equation} \hat{y}_{t+H} = y_{\,t+H-P}. \end{equation} 
For our minute-level Solana data, which exhibit strong intra-day periodicity, we set $P = 1440$ (minutes per day). Despite its simplicity, SNaive often provides a surprisingly competitive lower-bound benchmark in empirical forecasting studies\cite{hyndman2018forecasting}.

\subsection{Evaluation Metrics}

Mean Absolute Error (MAE) measures the average magnitude of errors between predictions \(\hat{y}_t\) and actual values \(y_t\):
\begin{equation}
MAE = \frac{1}{n} \sum_{t=1}^{n} \bigl|y_t - \hat{y}_t\bigr|,
\end{equation}
Root Mean Squared Error (RMSE) quantifies the square root of the average of squared differences between predictions and observations:
\begin{equation}
RMSE = \sqrt{\frac{1}{n} \sum_{t=1}^{n} \bigl(y_t - \hat{y}_t\bigr)^2},
\end{equation}
Mean Absolute Percentage Error (MAPE) expresses the average relative error without a percentage multiplier:
\begin{equation}
MAPE = \frac{1}{n} \sum_{t=1}^{n} \left\lvert\frac{y_t - \hat{y}_t}{y_t}\right\rvert.
\end{equation}

\subsection{Statistical Significance Tests}

Let $e_{i,j}$ denote the evaluation error (e.g., MAE/RMSE/MAPE; lower is better) of model $j\in\{1,\dots,k\}$ on dataset/series $i\in\{1,\dots,N\}$. We assess whether performance differences are statistically significant using a nonparametric omnibus test (Friedman), a paired post-hoc test (Wilcoxon signed-rank), and a parametric mixed-effects likelihood ratio test (MixedLM LRT).

\paragraph{Friedman test (omnibus, $k>2$ models).}
Within each dataset $i$, rank the $k$ models by ascending error to obtain $R_{i,j}$ (average ranks for ties). Let $R_j=\sum_{i=1}^{N} R_{i,j}$ be the sum of ranks and $\bar{R}_j=R_j/N$ the average rank. The Friedman statistic is
\begin{equation}
\chi_F^2 \;=\; \frac{12}{N\,k\,(k+1)}\sum_{j=1}^{k} R_j^2 \;-\; 3N\,(k+1)
\;=\; \frac{12N}{k\,(k+1)}\sum_{j=1}^{k} \bar{R}_j^2 \;-\; 3N\,(k+1).
\end{equation}
For moderate $N,k$, an improved finite-sample approximation uses the Iman--Davenport $F$ statistic:
\begin{equation}
F_F \;=\; \frac{(N-1)\,\chi_F^2}{N\,(k-1)-\chi_F^2} \;\sim\; F_{k-1,\,(k-1)(N-1)}.
\end{equation}
A significant result indicates at least one model differs; pairwise localization then proceeds with Wilcoxon (below), optionally with multiplicity control (e.g., Holm).

\paragraph{Wilcoxon signed-rank test (paired, post-hoc).}
For two models $A$ and $B$, form paired differences across datasets $d_i = e_{i,A}-e_{i,B}$. Discard zeros, let $n$ be the number of nonzero pairs, and rank the absolute differences $|d_i|$ to obtain ranks $r_i$ (average ranks for ties). Define the signed-rank sums
\begin{equation}
W^{+}=\sum_{d_i>0} r_i, \qquad W^{-}=\sum_{d_i<0} r_i, \qquad T=\min\{W^{+},\,W^{-}\}.
\end{equation}
For large $n$, a normal approximation is
\begin{equation}
z \;=\; \frac{W^{+}-\frac{n(n+1)}{4}}{\sqrt{\,\frac{n(n+1)(2n+1)}{24}\,}},
\end{equation}
yielding a two-sided $p$-value (a continuity correction of $0.5$ can be applied in the numerator if desired). We report $p$-values for each pair and apply a familywise adjustment when testing multiple pairs.

\paragraph{Mixed-effects LRT (parametric, model effect under clustering).}
To account for between-series heterogeneity and repeated measures, we fit a linear mixed model to the per-series errors:
\begin{equation}
e_{i,j} \;=\; \mu \;+\; \alpha_j \;+\; b_i \;+\; \varepsilon_{i,j},
\end{equation}
where $\mu$ is a global intercept, $\alpha_j$ is a fixed effect for model $j$ (with $\sum_j \alpha_j=0$ for identifiability), $b_i \sim \mathcal{N}(0,\sigma_b^2)$ is a random intercept for series $i$, and $\varepsilon_{i,j}\sim\mathcal{N}(0,\sigma^2)$ is residual noise. We fit both the null model $H_0:\alpha_1=\cdots=\alpha_k=0$ and the alternative $H_1$ using \emph{maximum likelihood} (ML). Let $\ell_0$ and $\ell_1$ be the maximized log-likelihoods under $H_0$ and $H_1$, respectively. The likelihood ratio statistic is
\begin{equation}
\Lambda \;=\; 2\bigl(\ell_1 - \ell_0\bigr) \;\sim\; \chi^2_{\,k-1},
\end{equation}
which tests for any overall performance effect across models while controlling for series-level clustering via $b_i$. When the distribution of $e_{i,j}$ is skewed, we analyze a monotone transform (e.g., $\log(e_{i,j}+\epsilon)$) and check residual diagnostics.

\section{Dataset}

\subsection{Dataset Description}

\begin{table}[!htbp]
\centering
\caption{Dataset description}
\label{tab:dataset_description}
\begin{tabular}{@{}lp{8.5cm}@{}}
\toprule
\textbf{Column Name} & \textbf{Description} \\
\midrule
token\_address & Address of the Solana token (1584 tokens) \\
daily\_date & From 2024 March 24 to 2025 March 16th, in 1-day intervals. Different tokens have their creation date as the starting data point of the time series. The minimum number of days is 49 and the average number of days is 153. \\
daily\_transaction\_count & Number of transfers for a single token on that day. \\
daily\_active\_wallet\_count & Number of active wallets for a single token on that day. \\
avg\_transaction\_size & Average trading volume of a single token for the day. \\
max\_transaction\_size & Maximum trading volume of a single token for the day. \\
daily\_trade\_volume & Daily transaction volume for a single token. \\
daily\_open\_price & The opening price of a single token (first hourly average selected to avoid errors). \\
daily\_close\_price & The closing price of a single token (last hour average price selected to avoid errors). \\
daily\_high\_price & Highest price of the day for a single token. \\
daily\_low\_price & Lowest price of the day for a single token. \\
daily\_avg\_price & Average price of the day for a single token. \\
daily\_avg\_bid & Average bid price. \\
daily\_avg\_ask & Average ask price. \\
daily\_avg\_spread & Average spread price. \\
Sellers & Number of sellers. \\
Buyers & Number of buyers. \\
unique\_count\_trader & Number of unique traders for the day (de-duplicated). \\
remainder\_usd & Estimated remaining funds in the liquidity pool for the day (USD). \\
cum\_remainder\_usd & Estimated remaining funds in the total liquidity pool at that date (USD). \\
cum\_unique\_trader & Total number of unique traders. \\
market\_value & Market value (price multiplied by total supply). \\
new\_traders & Number of new traders on the day. \\
solana\_price & Solana price (USD). \\
ma\_50 & Solana price 50-day Moving Average. \\
ma\_100 & Solana price 100-day Moving Average. \\
ma\_200 & Solana price 200-day Moving Average. \\
amount\_usd & Total DEX trading volume for the day (USD). \\
Traders & Total number of DEX traders for the day. \\
daily\_pairs & Number of new pairs added to DEX on that day. \\
\bottomrule
\end{tabular}
\end{table}

The dataset for this study comprises Solana tokens and was compiled using data sourced from DEX and the Solana blockchain. The dataset includes 27 daily features for each token. To ensure the quality and relevance of the data, a two-stage refinement process was applied. First, tokens were filtered using three activity and liquidity criteria: at least 2,000 unique holders, an average daily transaction count above 1,000, and total trading volume above 20 million USD. Second, subsequent to this initial filtering, a data cleaning procedure was carried out to remove tokens with extensive missing values, extreme abnormal deviations, or insufficient continuous observations for the forecasting setup. Table 1 details the composition of the dataset for this study.

\subsection{Data analysis}

\begin{figure}[!htbp]
  \centering
  \begin{minipage}[t]{0.48\textwidth}
    \centering
    \includegraphics[width=\textwidth]{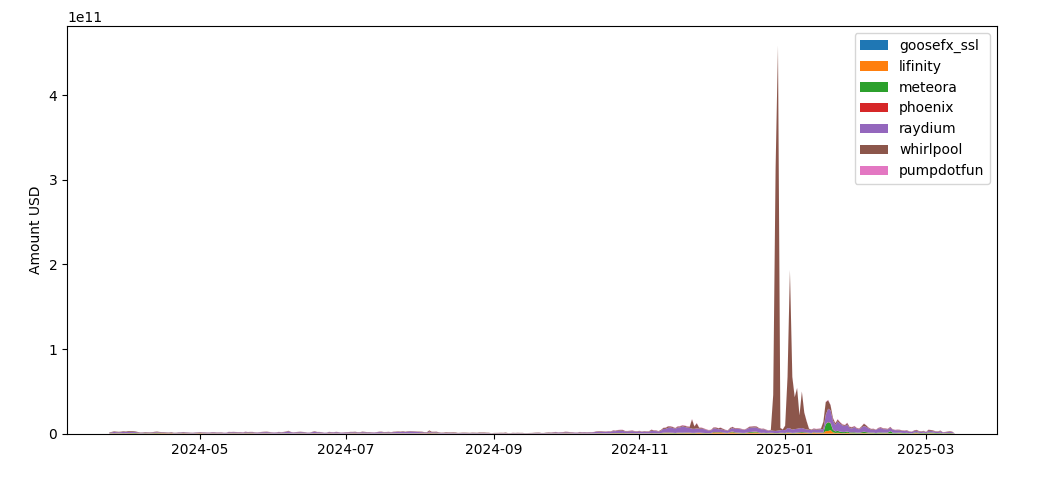}\\
    (a) Solana DEX Trading Volume
    \\[1em]
    \includegraphics[width=\textwidth]{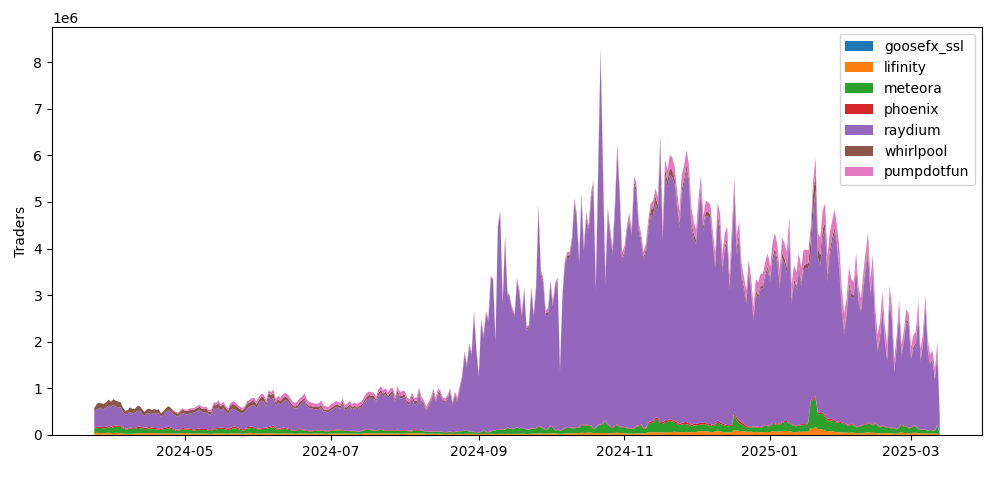}\\
    (b) Solana DEX Traders
    \\[1em]
    \includegraphics[width=\textwidth]{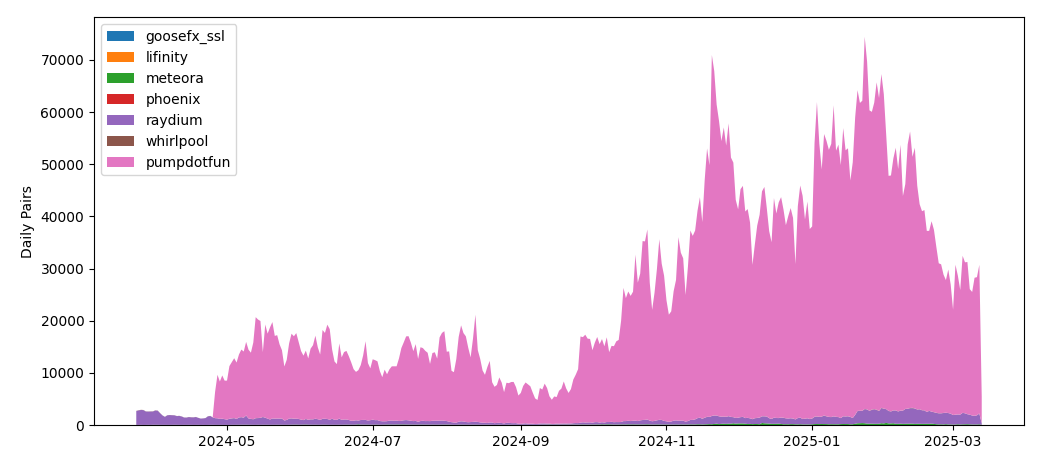}\\
    (c) New pairs added in Solana DEX
  \end{minipage}
  \hfill
  \begin{minipage}[t]{0.48\textwidth}
    \centering
    \includegraphics[width=\textwidth]{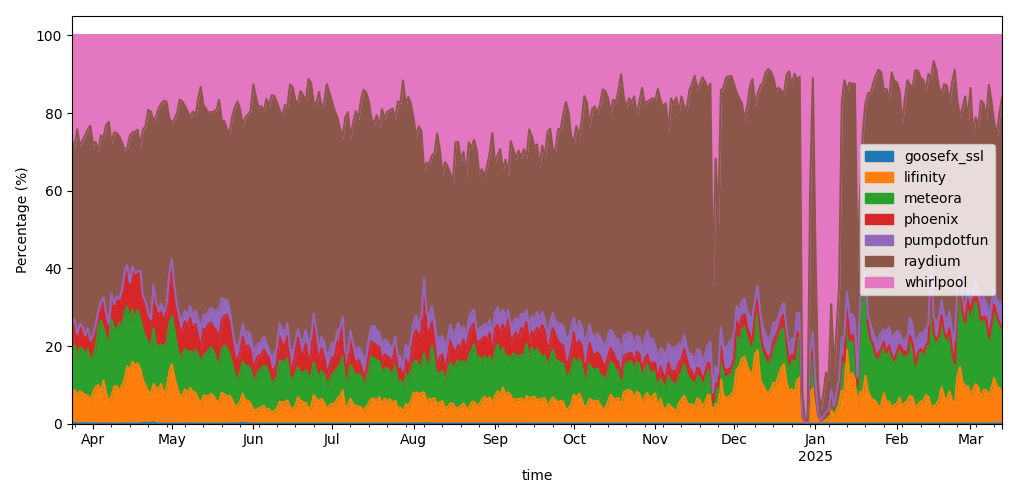}\\
    (d) Solana DEX Volume Share
    \\[1em]
    \includegraphics[width=\textwidth]{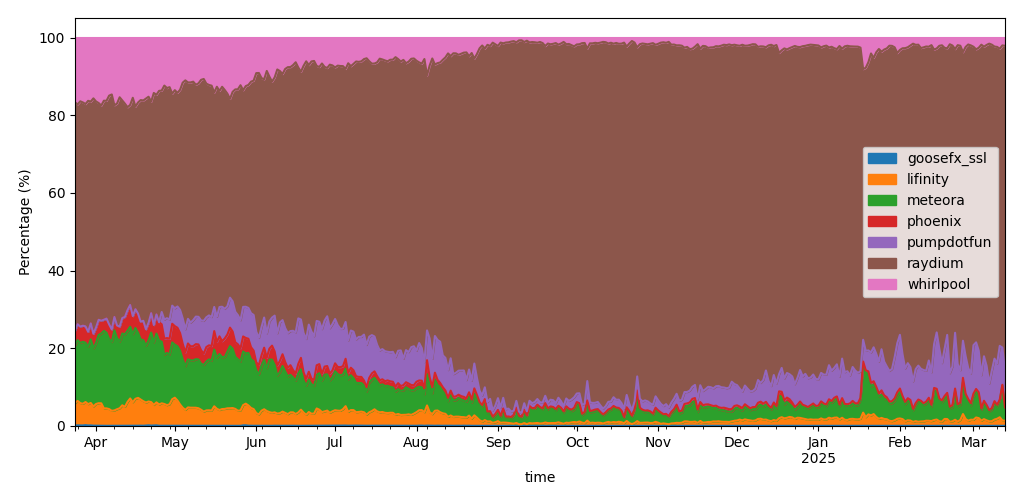}\\
    (e) Solana DEX Trader Share
    \\[1em]
    \includegraphics[width=\textwidth]{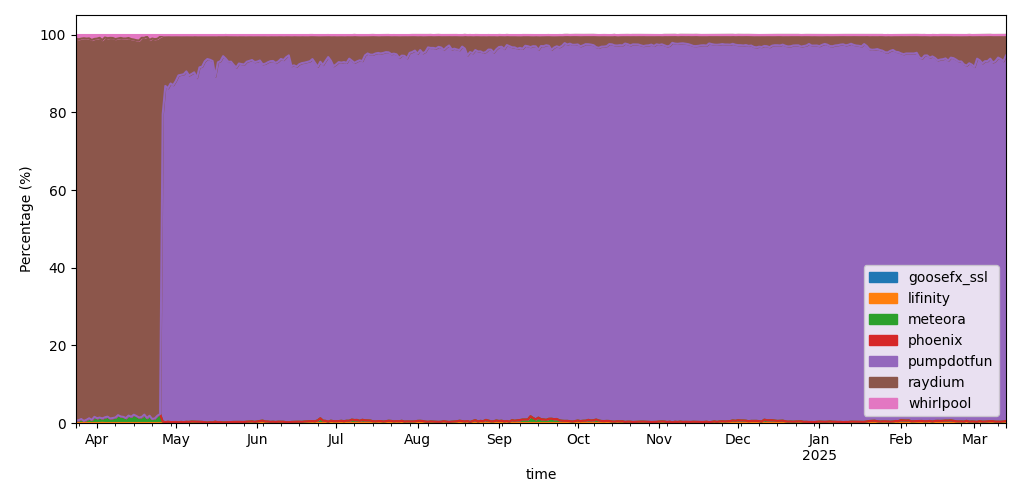}\\
    (f) Solana DEX New Pair Share
  \end{minipage}
  \caption{Solana DEX Metrics and Share}
  \label{fig:solana_dex_metrics}
\end{figure}

The main DEX on Solana are GooseFX, Lfinity, Meteora, Phoenix, Raydium, Whirlpool, and Pumpdotfun. Figure 1 shows that Raydium has consistently dominated in both trading volume and the number of traders, followed by Whirlpool, Lfinity, Meteora, Phoenix, and Pumpdotfun. GooseFX holds the smallest share. Notably, while Whirlpool has fewer traders than Raydium, it has consistently maintained the second-highest trading volume, reaching its peak in January 2025 when it accounted for the majority of the trading volume. Before the end of April 2024, Raydium had launched the most new trading pairs and held the largest share of this activity. After the end of April 2024, Pumpdotfun consistently launched the most new trading pairs, likely due to the ease of use of its Solana protocol. Observing the charts, there were two peaks at the end of 2024. The second peak in trading volume occurred shortly after a social media post on platform X by 47th U.S. President Trump on January 17, 2025, referencing the token 'Trump'.

\begin{figure}[!htbp]
  \centering
  \begin{minipage}[t]{0.3\textwidth}
    \centering
    \includegraphics[width=\textwidth]{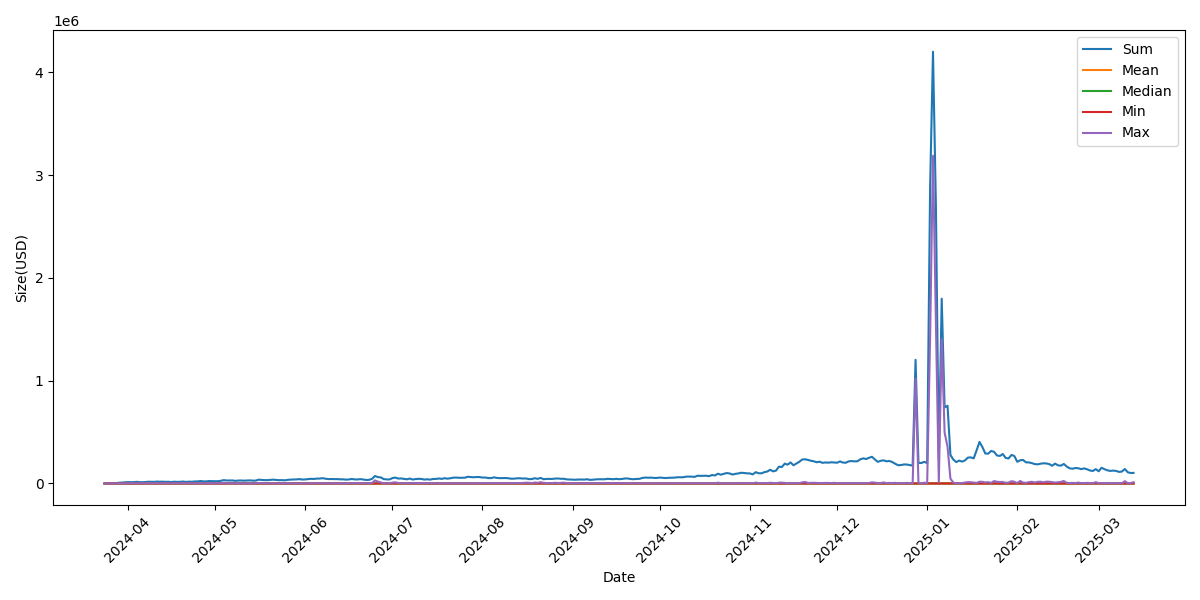}\\
    (a) Daily Average Transaction Size Trend(mean values)
  \end{minipage}
  \hfill
  \begin{minipage}[t]{0.3\textwidth}
    \centering
    \includegraphics[width=\textwidth]{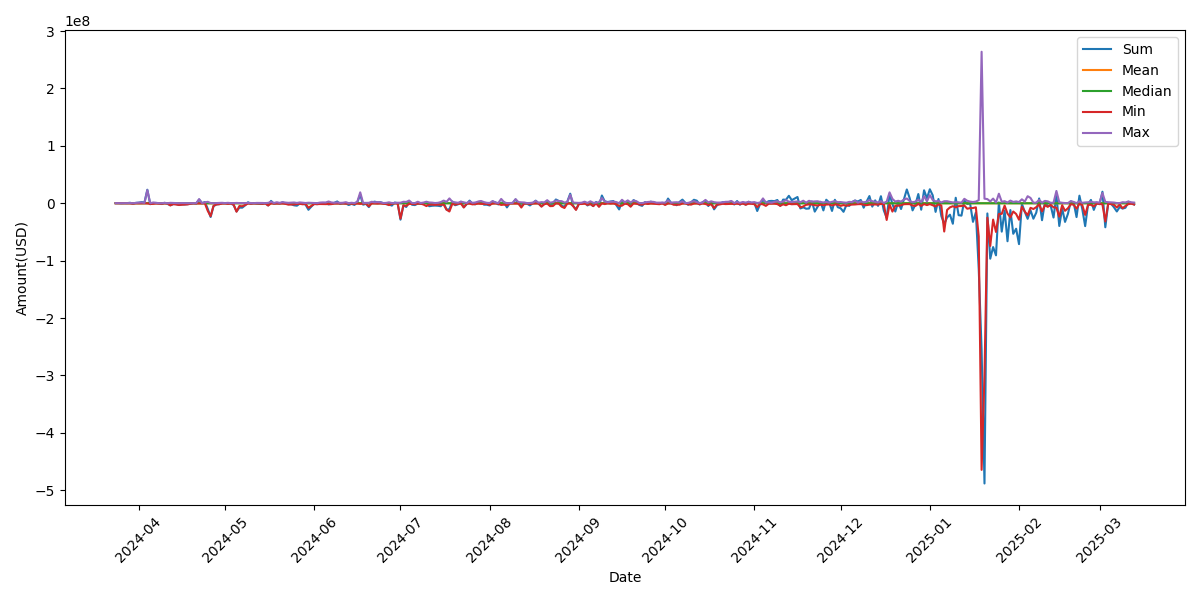}\\
    (b) Daily Remaining Funds in Liquidity Pool(mean values)
  \end{minipage}
  \hfill
  \begin{minipage}[t]{0.3\textwidth}
    \centering
    \includegraphics[width=\textwidth]{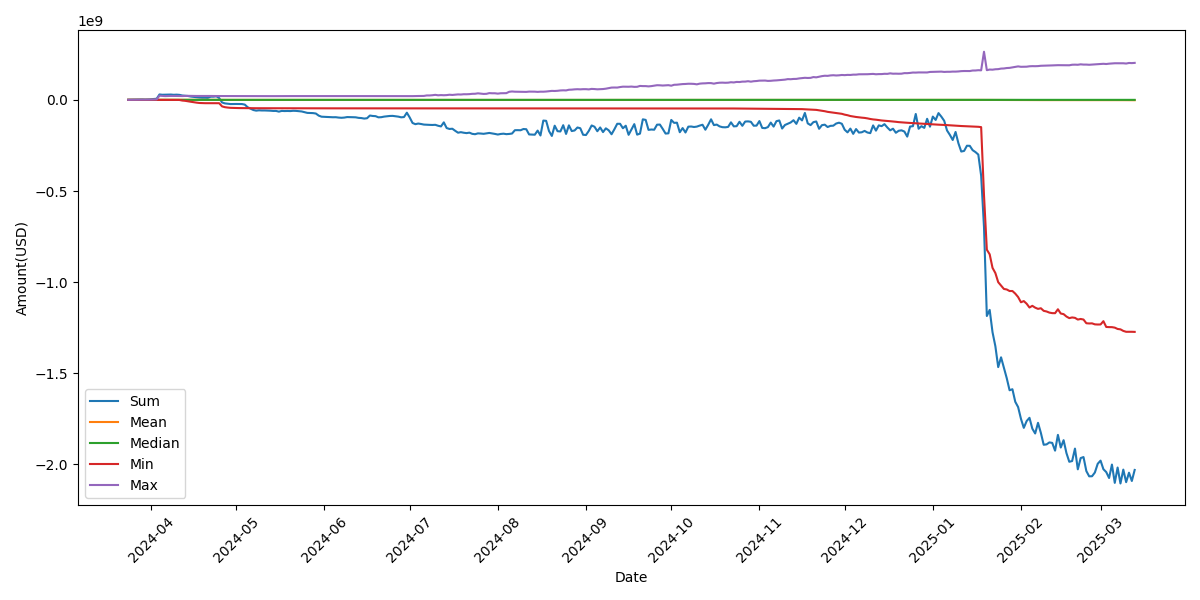}\\
    (c) Cumulative Remaining Funds in Liquidity Pool(mean values)
  \end{minipage}
  \caption{Token Transaction Size and Liquidity Pool}
  \label{fig:token_transaction_liquidity}
\end{figure}

Figure 2 shows that the peak in token transactions within the dataset occurred around January 2025. Examining the average liquidity pool of the dataset tokens reveals a pattern of increased investment leading up to mid-January 2025, followed by a period of increased withdrawal. We interpret this trend to suggest a shift in investor behavior, with a tendency towards increased investment before mid-January 2025 transitioning to a propensity for profit-taking thereafter. This inflection point coincides with the release of the token 'Trump'.

\begin{figure}[!htbp]
  \centering
  \begin{minipage}[t]{0.48\textwidth}
    \centering
    \includegraphics[width=\textwidth]{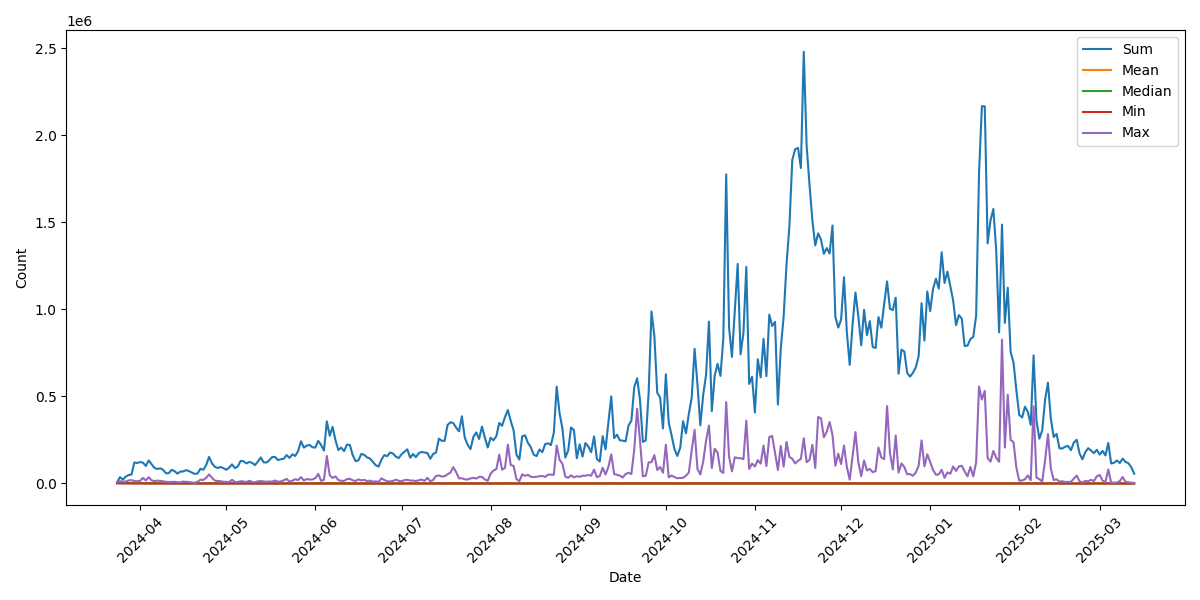}\\
    (a) Daily Active Wallet Count Trend(sum values)
    \\[1em]
    \includegraphics[width=\textwidth]{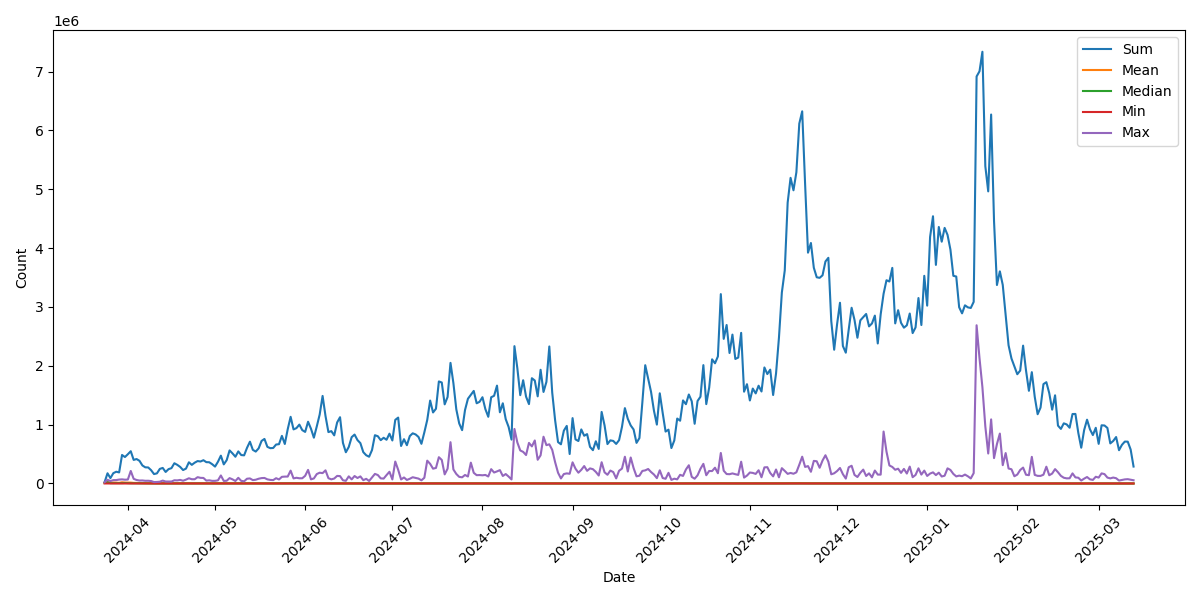}\\
    (b) Daily Transactions Count Trend(sum values) 
    \\[1em]
    \includegraphics[width=\textwidth]{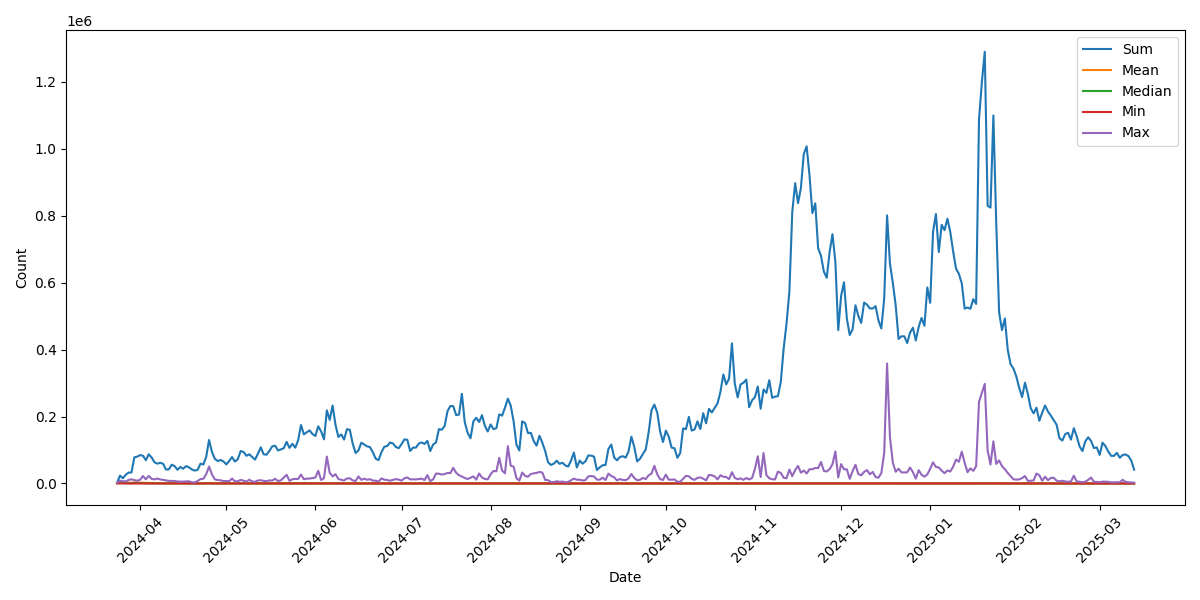}\\
    (c) Daily Number of Sellers Trend(sum values)
  \end{minipage}
  \hfill
  \begin{minipage}[t]{0.48\textwidth}
    \centering
    \includegraphics[width=\textwidth]{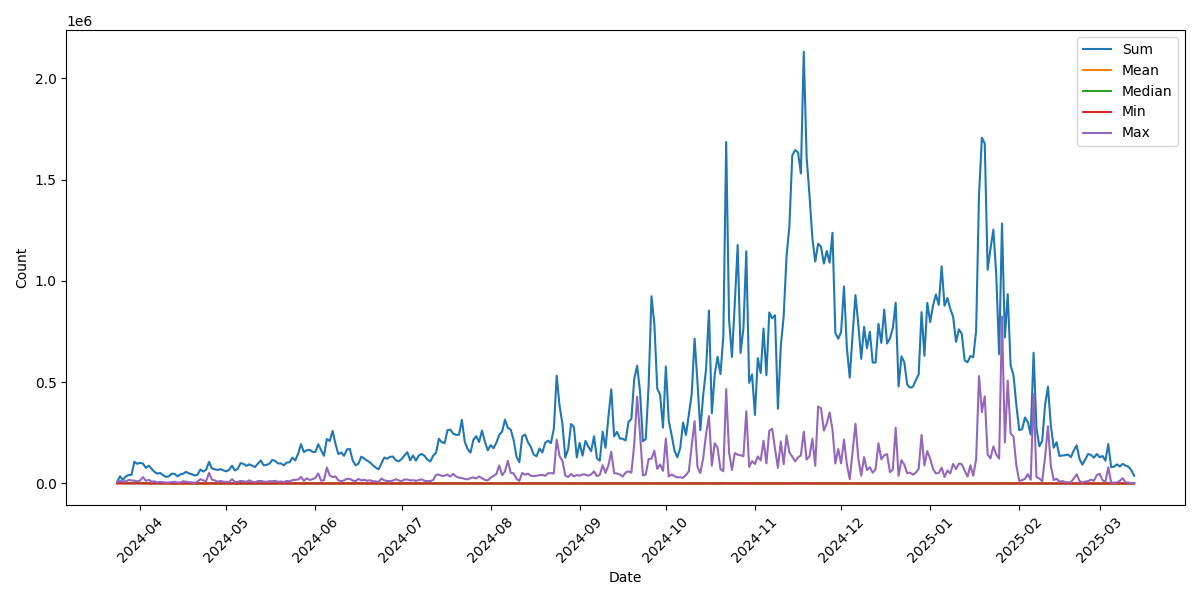}\\
    (d) Daily Number of Buyers Trend(sum values)
    \\[1em]
    \includegraphics[width=\textwidth]{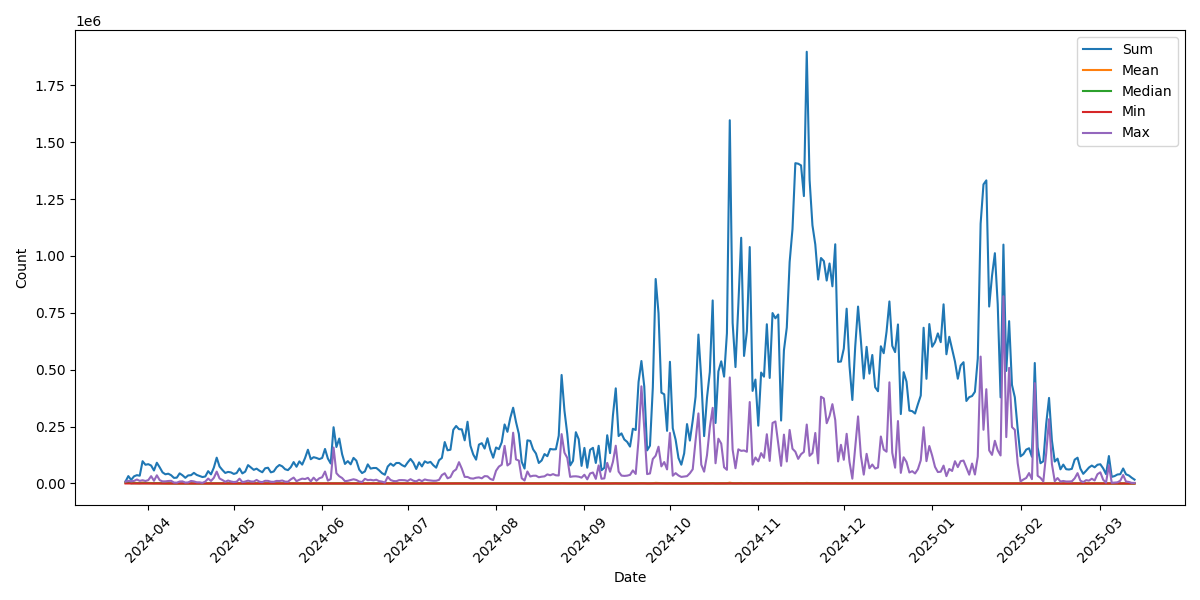}\\
    (e) Daily New Traders Count Trend(sum values)
    \\[1em]
    \includegraphics[width=\textwidth]{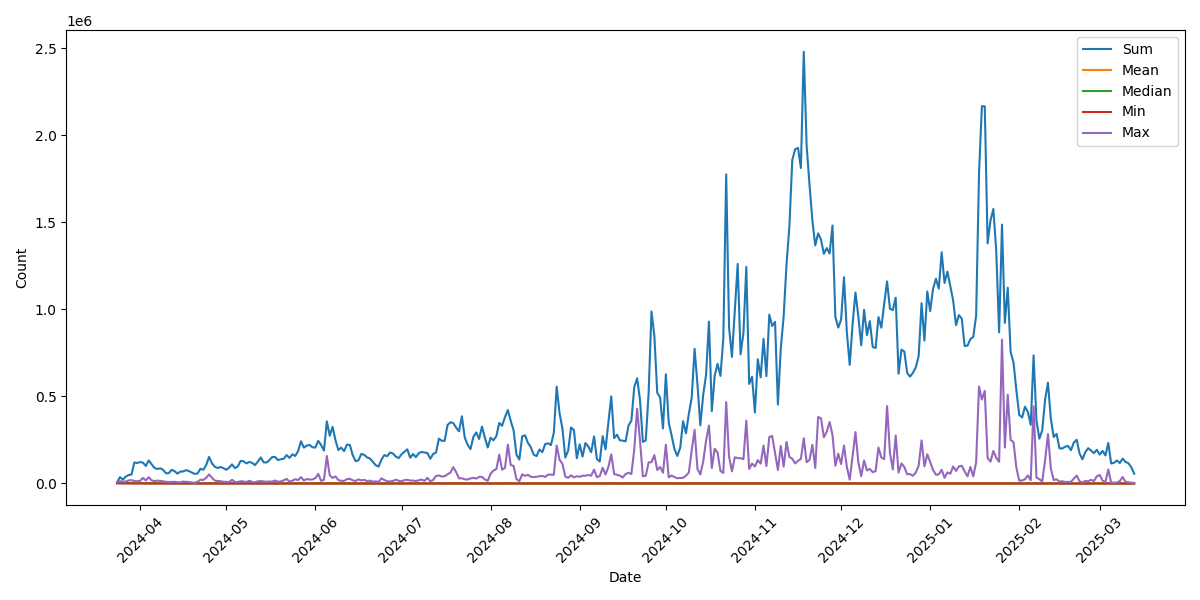}\\
    (f) Daily Unique Traders Count Trend(sum values)
  \end{minipage}
  \caption{Daily Activity Counts and Trends}
  \label{fig:daily_activity_counts}
\end{figure}

According to Figure 3, we observe two distinct peaks in several key on-chain metrics: the number of transfers, the number of sellers, the number of buyers, the number of new traders, the number of de-duplicated traders, and the number of active wallet addresses. These peaks occur around mid-November 2024 and mid-January 2025, respectively. Notably, these peaks show a strong temporal correlation with the two peaks in overall DEX trading volume presented in Figures 1-2. This alignment suggests a close relationship between user activity, trading behavior, and the overall dynamics of the Solana DEX market. While these correlations are evident, further analysis is needed to determine the underlying causal factors driving these synchronized fluctuations. The mid-January peak, in particular, coincides with the release of the 'Trump' token, as discussed previously, which likely contributed to the surge in these on-chain metrics.

\begin{figure}[!htbp]
  \centering
  \begin{minipage}[t]{0.3\textwidth}
    \centering
    \includegraphics[width=\textwidth]{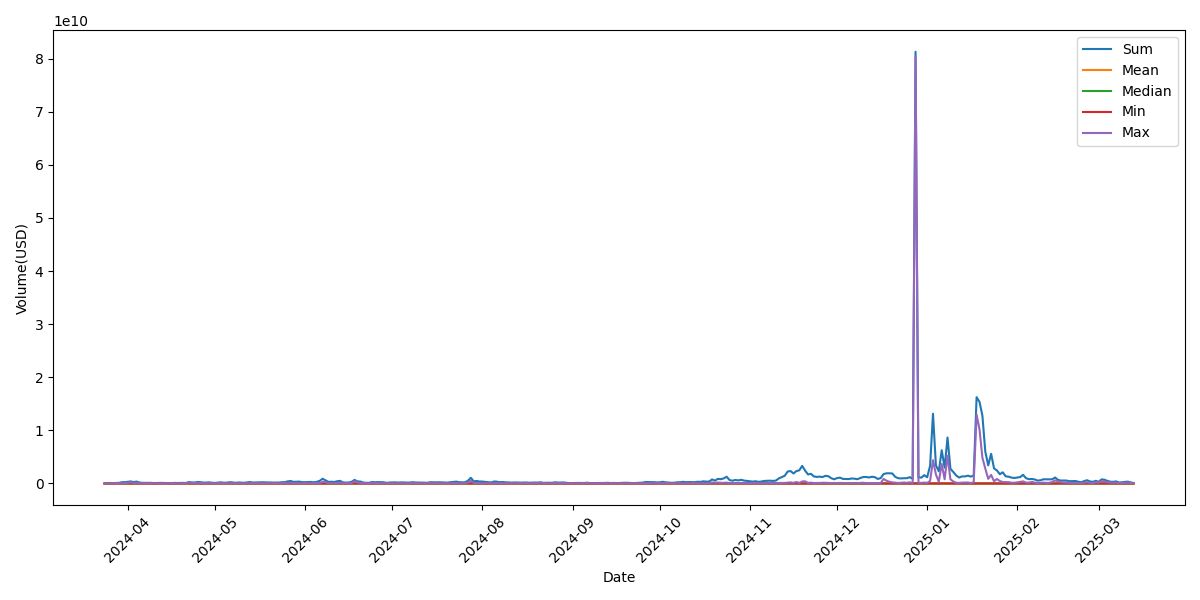}\\
    (a) Daily Trade Volume Trend (sum values)
  \end{minipage}
  \hfill
  \begin{minipage}[t]{0.3\textwidth}
    \centering
    \includegraphics[width=\textwidth]{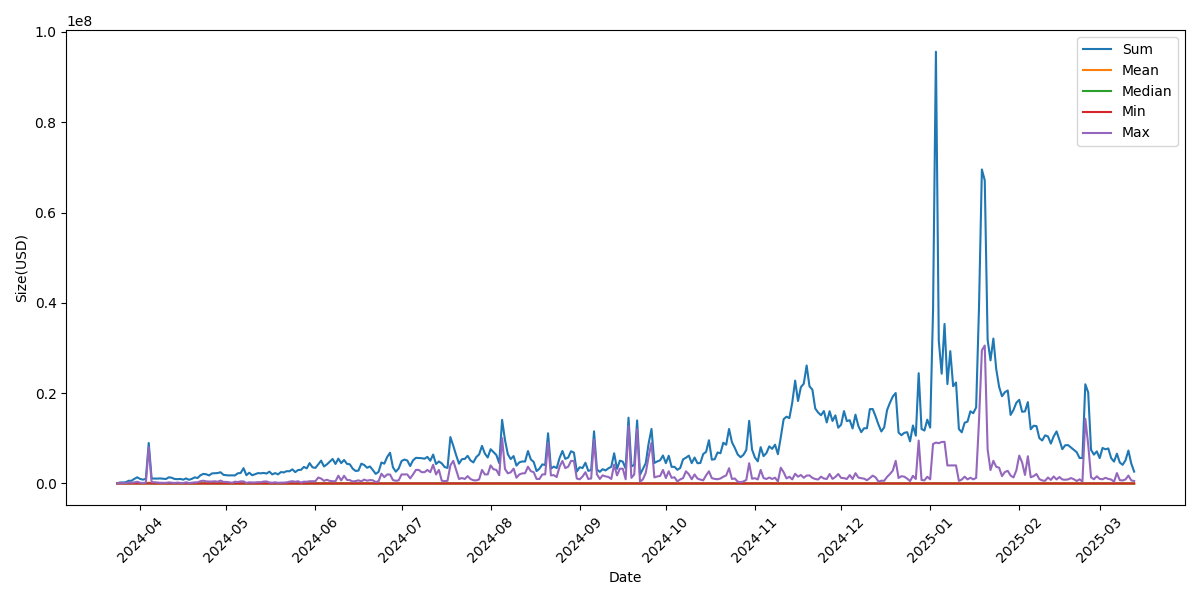}\\
    (b) Daily Max Transaction Size Trend(mean values)
  \end{minipage}
  \hfill
  \begin{minipage}[t]{0.3\textwidth}
    \centering
    \includegraphics[width=\textwidth]{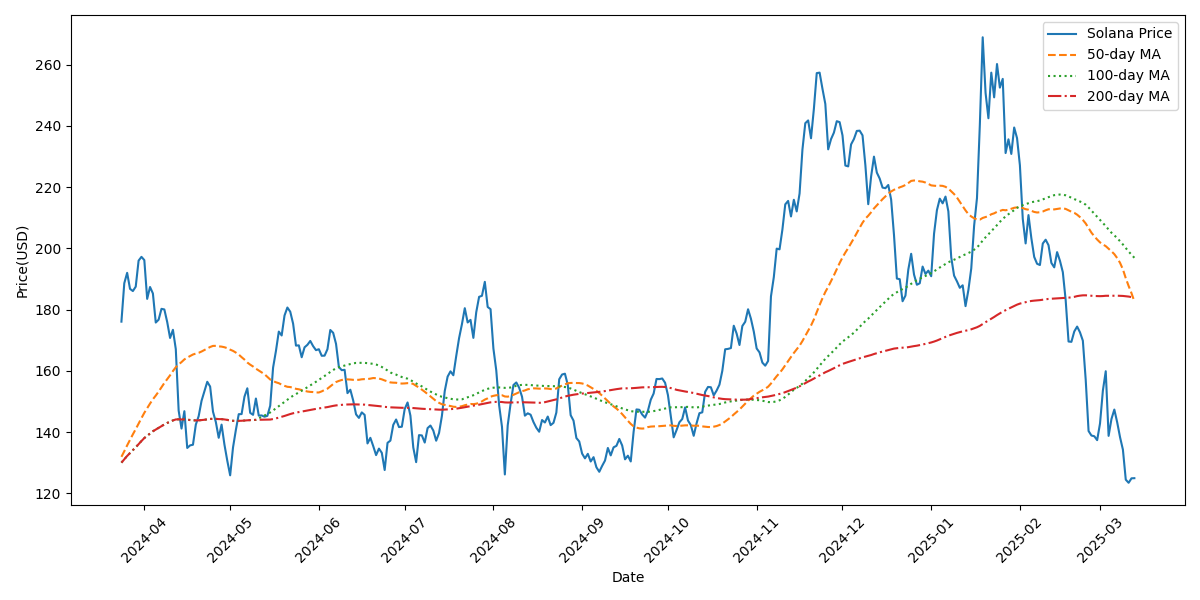}\\
    (c) Solana Price Indicators Trend Price(USD)
  \end{minipage}
  \caption{Token Trade Volume, Token Max Transaction Size and Solana Price}
  \label{fig:token_trade_volume_max_size_price}
\end{figure}

Figure 4(a) shows that the total daily average trading volume of the tokens in the dataset largely aligns with the overall Solana token DEX trading trend observed in Figure 1. Figure 4(b) reveals that while the period from November 2024 to February 2025 exhibited the highest trading volume, the number of trades experienced a significant peak in April 2024. According to Figure 4(c), Solana's price displayed two peaks in November 2024 and January 2025, generally matching the trading behavior of the tokens in the dataset and the peaks in the DEX market metrics presented in Figures 1 and 3.

\section{Results}

\subsection{Comparative Analysis}

We performed comparative experiments using the following models: PatchTST, TemporalFusionTransformer, TiDE, ChronosFineTuned [bolt\_small], ChronosZeroShot [bolt\_base], and TimeGPT. In addition to machine learning models and zero-shot time series forecasting models, we add additional statistical models based on seasonality and trends: SeasonalNaive and Prophet.

Experimental Setup Details:

\begin{enumerate}
\item The dataset contains 1584 tokens, which are time series of varying lengths. To ensure the robustness and temporal adaptability of the models, we employed a rolling window backtesting strategy on the historical data for validation. The final reported metrics, however, are based on a fixed, future-facing test set with a prediction length of $H=3$ days. This short horizon was specifically selected because assets in this ecosystem, particularly blockchain-based meme tokens, typically follow much shorter trading cycles than other conventional assets. Consequently, the 3-day forecast was chosen as a balanced, short-term prediction window, reflecting the practical limitations and time-sensitivity required for effective trading and risk management in this volatile ecosystem (particularly for assets with shorter trading cycles, such as meme tokens). This prediction length also helped mitigate certain uncontrollable factors inherent in comparing diverse models on highly volatile data. The final out-of-sample predictions, which serve as the basis for our reported metrics, are specifically for the three days from March 14th to 16th, 2025. We acknowledge, however, that the inherent volatility of cryptocurrencies might lead to very sparse data for certain tokens during this final prediction window.
\item Regarding the application of the models in our experiments, we maintained a crucial distinction: Deep learning models (PatchTST, TFT, TiDE) were trained and applied in a univariate manner, meaning each model was independently trained on a single token's time series. This approach resulted in 1584 separate trained models and 1584 training/prediction runs for the DL methods. Conversely, the zero-shot models (Chronos, TimeGPT) and the fine-tuned Chronos variant were applied in a multivariate fashion (or zero-shot transfer), where a single model was used to provide predictions across all 1584 tokens simultaneously.
\item We forecast token market capitalization (price $\times$ circulating supply) rather than raw price to normalize heterogeneous supply dynamics (issuance, burns, unlocks), align the target with ecosystem-level capital flows, and reduce sensitivity to thin-liquidity DEX dislocations.
\item The evaluation metrics used are MAE, RMSE, and MAPE. Regarding the second point, predicting price instead of market capitalization would introduce significant deviations in the MAE and RMSE scores when comparing different models because the total supply of different tokens varies. This would, for example, prevent the average of the evaluation metrics from accurately reflecting the prediction level of the corresponding model.
\item Data partitioning into training, validation, and test sets employed a time-ordered evaluation protocol and rolling window backtesting to ensure reliable out-of-sample performance and prevent look-ahead bias. For models managed by AutoGluon (PatchTST, TFT, TiDE, and Chronos), the presets='best\_quality' configuration was used, leveraging the framework's most powerful ensemble and automated tuning strategies. TimeGPT and Prophet were executed using their standard, pre-set prediction configurations.
\item var is the variance, median is the median, and RO represents the quantile mean calculated at intervals from 0.1 to 0.9. 
\item The experiments were conducted using the Google Colaboratory (Colab) platform, utilizing an T4 GPU.
\end{enumerate}

\begin{table}[!htbp]
  \centering
  \caption{Comparative Results Across Models and Metrics (MAE, RMSE, MAPE)}
  \label{tab:merged_results}
  \setlength{\tabcolsep}{1pt}%
  \fontsize{6}{7}\selectfont
  \begin{tabular}{@{}l rrr rrr rrr@{}}
    \toprule
    \textbf{Model} & \multicolumn{3}{c}{\textbf{MAE}} & \multicolumn{3}{c}{\textbf{RMSE}} & \multicolumn{3}{c}{\textbf{MAPE}} \\
    \cmidrule(lr){2-4} \cmidrule(lr){5-7} \cmidrule(lr){8-10}
    & $var$ & $median$ & $RO$ & $var$ & $median$ & $RO$ & $var$ & $median$ & $RO$ \\
    \midrule
    ChronosFineTuned & $7.258{\times}10^{20}$ & $9352.679$ & $43889.510$ & $7.314{\times}10^{20}$ & $9970.350$ & $46893.034$ & $5.050{\times}10^{8}$ & $0.181$ & $0.238$ \\
    ChronosZeroShot & $2.170{\times}10^{19}$ & $10322.112$ & $\textbf{31420.909}$ & $3.242{\times}10^{19}$ & $10982.573$ & $\textbf{33891.518}$ & $4.275{\times}10^{7}$ & $0.194$ & $0.268$ \\
    PatchTST & $7.846{\times}10^{20}$ & $6126.644$ & $54979.815$ & $7.928{\times}10^{20}$ & $6914.969$ & $60497.878$ & $1.087{\times}10^{7}$ & $0.143$ & $\textbf{0.223}$ \\
    TFT & $5.257{\times}10^{20}$ & $8987.153$ & $55297.827$ & $5.336{\times}10^{20}$ & $9417.512$ & $59082.527$ & $2.541{\times}10^{8}$ & $0.177$ & $0.274$ \\
    TiDE & $3.910{\times}10^{20}$ & $9890.421$ & $80114.104$ & $4.025{\times}10^{20}$ & $10862.607$ & $90963.842$ & $3.137{\times}10^{8}$ & $0.200$ & $0.331$ \\
    TimeGPT & $8.360{\times}10^{19}$ & $20404.412$ & $60718.034$ & $1.378{\times}10^{20}$ & $21789.330$ & $68292.093$ & $1.234{\times}10^{-1}$ & $0.555$ & $0.579$ \\
    \bottomrule
  \end{tabular}
\end{table}

For the model results, we prioritize MAPE due to its ability to normalize scale across datasets, although MAE and RMSE remain important references. Based on the results presented in Tables 2, PatchTST demonstrates the best overall performance. Notably, the results of the fine-tuned Chronos model, ChronosFineTuned, are very close to those achieved by PatchTST. While ChronosFineTuned generally outperforms ChronosZeroShot and exhibits a larger variance (suggesting increased instability after fine-tuning), its performance level approaches that of the top-performing PatchTST after fine-tuning. TimeGPT, despite its overall poorer performance, shows the lowest variance in dataset result stability. This is particularly noteworthy considering TimeGPT's inherently low variance when not fine-tuned.

\begin{table}[!htbp]
  \centering
  \caption{MAE, RMSE, and MAPE Results for Prophet and SeasonalNaive Models}
  \label{tab:statistical_models_full_results}
  \begin{tabular}{@{}lrr@{}}
    \toprule
    \textbf{Metric} & \textbf{Prophet} & \textbf{SeasonalNaive} \\
    \midrule
    var\_mae & $1.050 \times 10^{18}$ & $5.562 \times 10^{20}$ \\
    median\_mae & $181.455$ & $5188.740$ \\
    RO\_mae & $443.968$ & $31073.874$ \\
    var\_rmse & $1.053 \times 10^{18}$ & $6.201 \times 10^{20}$ \\
    median\_rmse & $208.039$ & $5608.094$ \\
    RO\_rmse & $494.764$ & $34351.731$ \\
    var\_mape & $9.860 \times 10^{7}$ & $1.141 \times 10^{3}$ \\
    median\_mape & $0.004$ & $0.152$ \\
    RO\_mape & $0.008$ & $0.163$ \\
    \bottomrule
  \end{tabular}
\end{table}

\begin{table}[!htbp]
  \centering
  \caption{Model Comparison: Number of Tokens Outperforming Baselines}
  \label{tab:outperforming_baselines}
  \begin{tabular}{@{}lrr@{}}
    \toprule
    \textbf{Model} & \textbf{better\_than\_Prophet} & \textbf{better\_than\_SeasonalNaive} \\
    \midrule
    ChronosFineTuned & 33 & 683 \\
    ChronosZeroShot & 18 & 681 \\
    PatchTST & 93 & 772 \\
    TFT & 77 & 686 \\
    TiDE & 58 & 583 \\
    TimeGPT & 28 & 244 \\
    \bottomrule
  \end{tabular}
\end{table}

According to Table 3, we find that the results of Prophet and SeasonalNaive are much better than those of the other models in Tables 2, with Prophet performing particularly well. However, this does not necessarily mean that Prophet and SeasonalNaive are inherently superior to the other models; it suggests that statistical models relying on periodicity and trend perform better for most tokens within the prediction intervals considered in Table 3. Therefore, we have added Table 4 to analyze the number of other models with better MAPE results than Prophet and SeasonalNaive. We find that PatchTST performs the best in this regard, outperforming Prophet on 93 tokens and SeasonalNaive on 772 tokens. This indicates that PatchTST captures significant non-periodic and trending features that Prophet and SeasonalNaive miss. ChronosFineTuned outperforms SeasonalNaive on 683 tokens and Prophet on 33 tokens, while ChronosZeroShot outperforms SeasonalNaive on 681 tokens and Prophet on 18 tokens. This suggests that after fine-tuning, Chronos can understand the properties of non-periodicity and trends significantly better. In terms of the number of outperformances over Prophet, Chronos is closer to the deep learning models (though still less), but it is closer to the deep learning models in the number of outperformances over SeasonalNaive. TimeGPT, on the other hand, has the fewest outperformances over both Prophet and SeasonalNaive, suggesting that while TimeGPT may perform more consistently, it captures the least additional non-periodic and trending features.

\begin{table}[!htbp]
  \centering
  \setlength{\tabcolsep}{4pt}
  \caption{Average Ranks of Forecasting Models}
  \label{tab:model_ranking_updated}
  \begin{tabular}{@{}lcccc@{}}
    \toprule
    \textbf{Model} & \textbf{rank\_mae} & \textbf{rank\_rmse} & \textbf{rank\_mape} & \textbf{mean\_rank} \\
    \midrule
    PatchTST & \textbf{2.958333} & \textbf{2.988005} & \textbf{2.958965} & \textbf{2.968434} \\
    ChronosFineTuned & 3.277778 & 3.231061 & 3.270202 & 3.259680 \\
    ChronosZeroShot & 3.361111 & 3.295455 & 3.362374 & 3.339646 \\
    TemporalFusionTransformer & 3.403409 & 3.368687 & 3.410354 & 3.394150 \\
    TiDE & 3.891414 & 3.940657 & 3.878157 & 3.903409 \\
    TimeGPT & 4.802399 & 4.889520 & 4.799242 & 4.830387 \\
    NPTS & 6.305556 & 6.286616 & 6.320707 & 6.304293 \\
    \bottomrule
  \end{tabular}
\end{table}

According to Table 5, and in order to further rigorously compare the performance of the forecasting models, we adopted a statistical framework combining non-parametric (Friedman, Wilcoxon) and parametric (MixedLM LRT) tests. For global significance, the Friedman test (treating tokens as blocks) and a mixed-effects linear model with token random intercepts and model as a fixed effect (evaluated via a likelihood-ratio test) both yielded $p$-values $< 0.05$, indicating significant overall differences in model performance. As the primary performance indicator, we computed the average rank (mean\_rank): models were ranked in ascending order of error for each metric (MAE, RMSE, MAPE) on every token, and then averaged across metrics (where a lower rank is better). The results show that PatchTST achieved the lowest mean rank ($\approx 2.97$), followed by Chronos FineTuned ($\approx 3.26$) and ChronosZeroShot ($\approx 3.34$). Notably, ChronosFineTuned is significantly superior to ChronosZeroShot, clearly demonstrating the substantial performance uplift achieved through fine-tuning. In conclusion, these results confirm PatchTST's superior overall robustness, evidenced by its lowest average rank and highest number of significant wins.

\subsection{Feature Importance Analysis}

In order to analyze in further detail the importance of each feature for prediction and to better understand how the model understands the dataset, we performed a feature importance analysis for TFT and TimeGPT.

For TFT, feature importance is calculated by replacing each feature with a permuted version of the same feature and evaluating the relative degradation of the model's prediction performance.

For TimeGPT, feature importance is calculated using SHAP (SHapley Additive exPlanation) values. SHAP values apply game-theoretic concepts to explain how each feature affects the machine learning prediction\cite{lundberg2017unified}.

\begin{figure}[!htbp]
  \centering
  \includegraphics[width=0.45\textwidth]{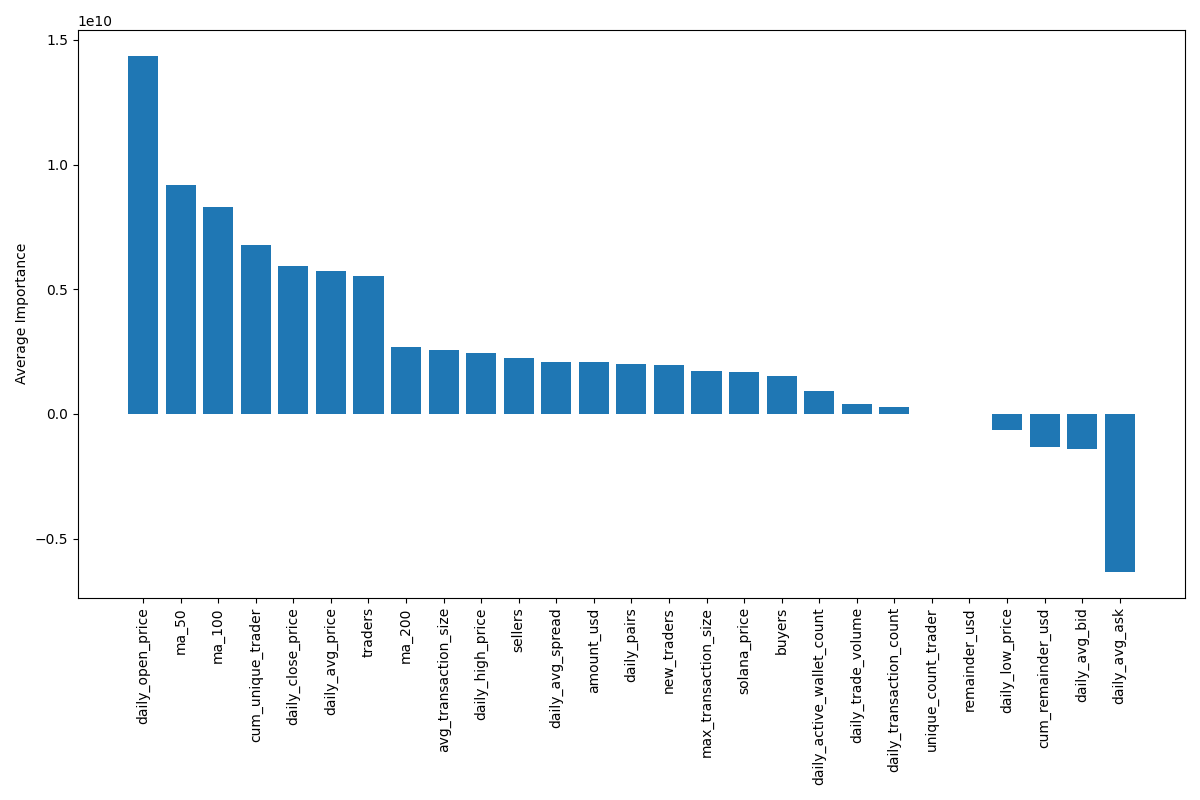}
  \caption{TFT Feature Importance}
  \label{fig:pic1_description} 
\end{figure}

Figure 5 indicates that the opening price of the token is the most important feature for TFT, followed by the 50-day and 100-day moving averages (MA50 and MA100) of the Solana price, the number of unique traders, and the closing price. Conversely, ask and bid prices appear to be the least important and have a negative influence.

\begin{figure}[!htbp]
  \centering
  \begin{minipage}[t]{0.45\textwidth}
    \centering
    \includegraphics[width=\textwidth]{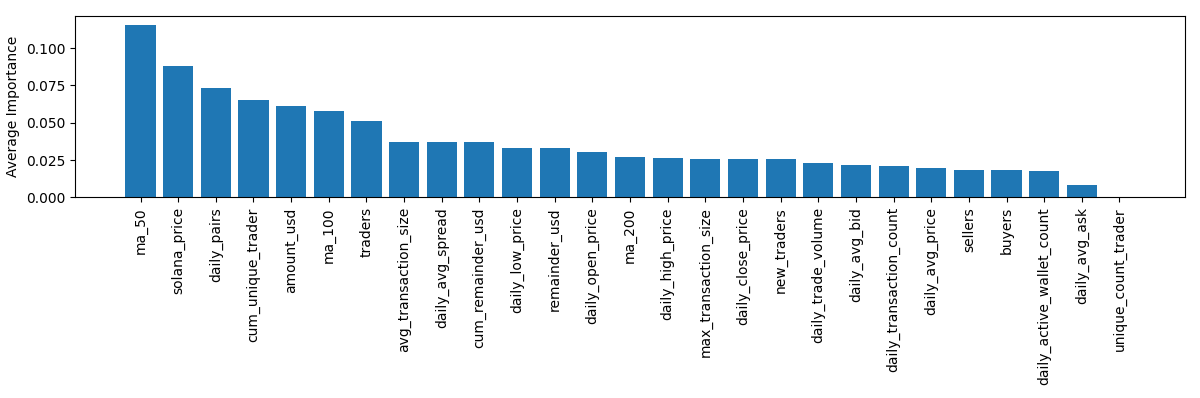}\\
    (a) Feature Importance ratios
  \end{minipage}
  \hfill
  \begin{minipage}[t]{0.45\textwidth}
    \centering
    \includegraphics[width=\textwidth]{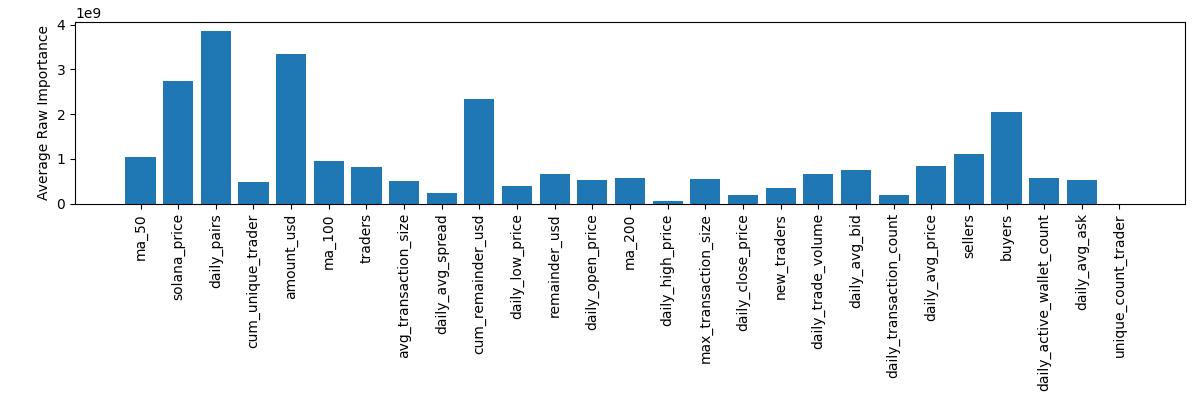}\\
    (b) Feature Importance
  \end{minipage}
  \caption{TimeGPT Feature Importance}
  \label{fig:two_pics_inline}
\end{figure}

According to Figure 6, when analyzing the average importance ratio, we find that TimeGPT's predictions are more dependent on Solana price and its moving averages (MA). Additionally, the daily generated pairs, the number of unique DEX traders (de-weighted), and market volume, which are characteristics derived from overall DEX data, are identified as more important features for TimeGPT.

For the Importance Mean analysis in Figure 6, Solana price, the number of DEX pairs traded daily, DEX market volume, liquidity pool size, and the number of purchasers are shown to be very important for predicting some specific tokens.

Combining the findings from Figures 5 and 6, it becomes evident that the overall numerical characteristics of the DEX market and the Solana price are the most important features affecting Solana token prices. This evidence indicates that the volatility of the DEX market and the Solana price are the key determinants of Solana token price movements.

\section{Discussion}

The central contribution of this study is the construction and analysis of a dedicated Solana digital asset time series dataset, rather than only a benchmark of forecasting algorithms. The dataset links token-level price, transaction, liquidity, and trader-activity variables with ecosystem-level Solana DEX indicators and SOL price features. This design makes it possible to study individual token behavior together with the broader DEX environment in which those tokens trade, which is especially important for assets with short histories, rapid listing cycles, and abrupt regime changes.

The descriptive analysis reveals that Solana token dynamics in this period were not isolated asset-level processes. We observed prominent activity peaks around mid-November 2024 and mid-January 2025, and these peaks appeared simultaneously in overall DEX volume, dataset token trading volume, active wallets, buyers, sellers, new traders, unique traders, liquidity-pool indicators, and SOL price. The January 2025 peak, which coincides with the 'Trump' token event, is particularly informative because the liquidity indicators show an accumulation phase before mid-January followed by increased withdrawals. This pattern suggests that the dataset captures a market-wide transition from inflow-driven activity to profit-taking behavior, although the evidence should be interpreted as temporal association rather than causal proof.

The forecasting benchmark serves as a second layer of dataset validation. By forecasting token market capitalization over a three-day horizon, we tested whether the constructed variables contain usable short-term predictive signal under the sparse and volatile conditions of Solana digital assets. PatchTST achieved the best overall average rank, and fine-tuned Chronos followed closely, indicating that the dataset supports models capable of capturing non-periodic token fluctuations. At the same time, Prophet and SeasonalNaive remained strong for many tokens, which shows that a substantial part of the dataset also contains short-term persistence, trend, or seasonal structure.

The feature-importance results further connect the forecasting task back to dataset interpretation. Token-specific price history remains important, but macro-level Solana variables--including SOL price, SOL moving averages, total DEX trading volume, DEX trader counts, newly created pairs, liquidity-pool measures, and buyer activity--consistently appear among the influential covariates. This indicates that individual token market capitalization is strongly tied to broader Solana DEX conditions, and that forecasting performance depends not only on local token history but also on ecosystem-level activity and sentiment.

In conclusion, this work contributes a curated dataset and an empirical analysis of Solana digital assets during a high-growth, high-volatility period. The dataset documents synchronized DEX, on-chain, liquidity, and price movements; highlights event-associated market shifts; and shows that macro Solana DEX indicators are central to explaining individual token volatility. The model comparison is therefore best understood as supporting evidence that the dataset contains both trend-like and non-periodic predictive structure. Future work should extend the temporal coverage, test forward-rolling vintages after March 2025, and replicate the same data design on other blockchain ecosystems to evaluate how much of the observed behavior is Solana-specific.

\bmhead{Acknowledgments}

Acknowledgements should be brief, and should not include thanks to anonymous referees and editors, or effusive comments. Grant or contribution numbers may be acknowledged.
\section*{Declarations}

\noindent\textbf{Funding} The work of Pavel Braslavski and Dmitry I. Ignatov was supported by the Basic Research Program at the National Research University, Higher School of Economics (HSE University).

\noindent\textbf{Author contributions} YX and MW contributed equally to this article as co-first authors. YX and MW wrote the manuscript, acquired and analyzed the data, and performed experiments. MW designed the study. PB and DI reviewed and corrected the article.

\noindent\textbf{Competing interests} The authors declare no competing interests.

\noindent\textbf{Correspondence} Correspondence and requests for materials should be addressed to M.W.

\bibliography{sn-bibliography}

\end{document}